\documentclass[pre,twocolumn,showpacs,floatfix,nofootinbib,groupaddress,showkeys]{revtex4-1}

\usepackage{amsmath}
\usepackage{amsfonts}
\usepackage{amssymb}
\usepackage[margin=0.75in]{geometry}
\usepackage{graphicx} 
\usepackage{cleveref}
\usepackage{bm}
\usepackage[outdir=./]{epstopdf}
\renewcommand{\[}{\begin{equation}}
\renewcommand{\]}{\end{equation}}

\begin{document}

\title{Undulation instabilities in cholesteric liquid crystals induced by anchoring transitions}
\author{Maxim O. Lavrentovich}
\email{lavrentm@gmail.com}
\affiliation{Department of Physics \& Astronomy, University of Tennessee, Knoxville, Tennessee 37996, USA}
\author{Lisa Tran}
\email{lt2727@columbia.edu}
\affiliation{Department of Chemical Engineering, Columbia University, New York, New York 10027, USA}

\begin{abstract}
Cholesteric liquid crystals (CLCs) have a characteristic length scale given by the pitch of the twisted stacking of their constituent rod-like molecules. Under homeotropic anchoring conditions where the molecules prefer to orient perpendicular to an interface, cholesteric interfaces  exhibit striped phases with stripe widths commensurate with the pitch. Conversely, planar anchoring conditions have the molecules remain in the plane of the interface so that the CLC twists perpendicular to it.  Recent work [L. Tran \textit{et al.} Phys. Rev. X \textbf{7}, 041029 (2017)]  shows that varying the anchoring conditions  dramatically rearranges the CLC stripe pattern, exchanging defects in the stripe pattern with defects in the molecular orientation of the liquid crystal molecules. We show with experiments and numerical simulations that the CLC stripes also undergo an undulation instability when we transition from homeotropic to planar anchoring conditions and vice versa. The undulation can be interpreted as a transient relaxation of the CLC resulting from a strain in the cholesteric layers due to a tilting pitch axis, with properties analogous to the classic Helfrich-Hurault instability. We focus on CLC shells in particular and show that the spherical topology of the shell also plays an important role in shaping the  undulations.  \end{abstract}
\maketitle


\section{Introduction}

\begin{figure}[htp]
\centering
\includegraphics[width=0.4\textwidth]{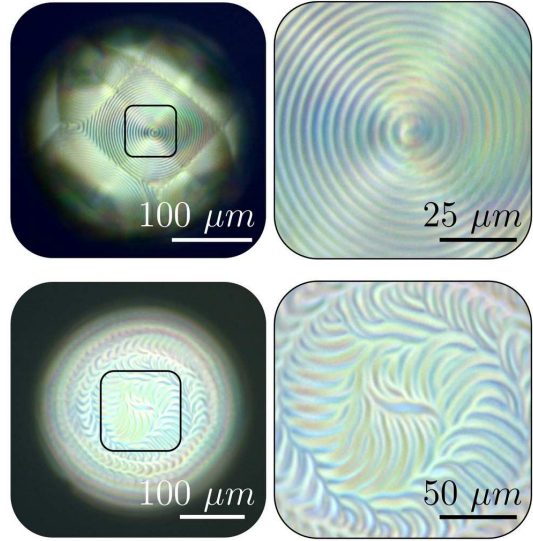}
\caption{\label{fig:FCDunwind}   A thick cholesteric shell with 1\% wt PVA, 7 mM SDS in the surrounding aqueous phases has homeotropic anchoring on the inner and outer interfaces. With 1 M NaCl in the inner phase and 0.1 M NaCl in the outer phase, the shell osmotically swells from top to bottom, diluting the surfactants in the inner phase and reducing the inner homeotropic anchoring strength. The planar stripes become thicker as a consequence and a secondary periodicity within primary stripes is visible, while maintaining the overall double spiral structure of a focal conic domain.}
\end{figure}

Striped patterns abound in nature, with lamellar features observable at the micron scale within the cell walls of fruits \cite{pollia}, the chitinous exoskeleton of beetles \cite{beetle}, and the fruit fly embryo \cite{fruitfly}, as well as at much larger scales, such as on the skin of zebras, tigers, and certain fish species \cite{fish,Turingfish}. In the latter examples, the stripes arise from activating and inhibiting dynamics, characteristic of a Turing instability \cite{fish,Turingfish}. Other patterns seen in living systems, such as those at the surfaces of \textit{chiral, liquid crystalline} materials, arise due to a characteristic length scale, \textit{e.g.}, the cholesteric pitch $P_0$ \cite{pollia, beetle, boulig-biochol, bouligarches}. Striped patterns can further be influenced by geometrical confinement. For instance, stripes decorating a sphere necessarily have defects where they must collide or terminate due to the system's topology, seen in the presence of poles on a globe. Motivated by the ubiquity of curved, chiral materials, we study striped patterns at the free surfaces of cholesteric liquid crystal (CLC) shells. Using varying surfactant concentrations in the ambient aqueous medium, we show that the CLC shell surface develops transient, undulated stripe instabilities as the pitch axis reorients to accommodate changes in the anchoring conditions. The instabilities are recapitulated in numerical simulations and are generic features: the transient undulations occur under either homeotropic to planar anchoring transitions or under the reverse change.  Fig.~\ref{fig:FCDunwind} illustrates this process for a cholesteric shell with diluting surfactant, triggering a homeotropic to planar anchoring transition. The initial striped, focal conic structure (double spiral, top row) unwinds and develops the secondary, bent stripes (bottom row). 

Recent work \cite{TranPRX} has shown that when cholesteric shells are subjected to modifications in the anchoring conditions, the stripe pattern is altered dramatically. When the anchoring transforms from homeotropic to planar, the stripes unwind at the defects. The topologically required defects in the system evolve from being defects in the twisting direction ($\lambda^{\pm}$ pitch defect lines) to becoming defects in the molecular orientation of the liquid crystalline molecules (nematic defects with a total charge of $+2$). Along with this conversion of the defect structure, the stripes themselves widen and develop an instability that generates a secondary periodicity, shown in the bottom row of Fig.~\ref{fig:FCDunwind}. Our objective here is to understand this instability using numerical simulations and to describe the stripe undulations through the lens of the Helfrich-Hurault instability. We show that the secondary stripe periodicity is caused by strain in the cholesteric layers brought about by changes in the anchoring condition.

The paper is organized as follows:  In the next section (Section~\ref{sec:setup}), we discuss the experimental setup and numerical simulations. Then, in Section~\ref{sec:Results}, we discuss our experimental and simulation results for both planar to homeotropic and homeotropic to planar anchoring transitions. In Section~\ref{sec:Analysis}, we present an analysis of the  undulating stripe patterns using the framework of the Helfrich-Hurault instability and predict a stripe periodicity with this model that matches those observed in experiments and simulations. We conclude with Section~\ref{sec:Conclusion} by giving perspectives for future work.


\section{Methods\label{sec:setup}}

\subsection{Cholesteric shell preparation}
    
We use 4-cyano-4'-pentylbiphenyl (5CB, Kingston Chemicals Limited) doped with 2.8\% wt (S)-4-cyano-4'-(2-methylbutyl)biphenyl (CB15, EMD Performance Materials) to obtain a CLC with a pitch of $\sim$ 5 $\mu$m. The pitch is determined using a Grandjean-Cano wedge cell \cite{GJwedge, Cwedge}. Briefly, two glass slides are coated with a unidirectionally rubbed polyvinyl alcohol (PVA, Sigma-Aldrich) film to give planar anchoring. The slides are then clamped and glued with rubbing directions parallel to one another and a Mylar spacer on one side of the cell to create the wedge angle. The distance between disclination lines in the cell is measured and used, along with the angle, to determine the pitch.
 
CLC shells are produced using a glass capillary microfluidic device to generate double emulsions, similar to methods described in past works \cite{ufluidics,TranPRX,tll-sds,lagerwall-review}. Briefly, three different fluid phases are used to generate shells of CLC: an inner aqueous phase, the middle CLC phase, and an outer aqueous phase. The tapered circular capillary of the inner water phase is fitted into a tapered square capillary for the middle CLC phase, which is subsequently fitted into a larger circular capillary for injection of the outer aqueous phase. The middle, square capillary is treated with octadecyltrimethoxysilane (OTS, Sigma-Aldrich) to improve CLC wetting of the capillary. Three syringe pumps (Harvard Apparatus) are used to control the flow rates of each phases, with 40 mL/hr, 500 $\mu$L/hr, and 400 $\mu$L/hr as the flow rates corresponding to the outer, middle and inner phases. Both aqueous phases have 1\% wt polyvinyl alcohol (PVA, 87-89\% hydrolyzed, average $M_{w} = 13,000-23,000$) to stabilize the emulsions. After the double emulsions are collected from the microfluidic device, the shells are left to settle in the vial and equilibrate to the planar anchoring conditions induced by PVA and water. 

We use sodium dodecyl sulfate as the surfactant (SDS, Sigma-Aldrich) in varying concentrations to induce homeotropic anchoring \cite{TranPRX,tll-sds,lagerwall-interface}. For homeotropic anchoring of the inner shell surface, SDS is included in the inner aqueous phase during the microfluidic production of CLC double emulsions. To induce homeotropic anchoring on the outer surface of the CLC shell, the double emulsions are pipetted into vials containing aqueous solutions of 1\% wt PVA, 0.1 M NaCl, and SDS with concentrations ranging from 7-10 mM \cite{TranPRX}. The shells are either viewed immediately to observe the planar to homeotropic anchoring transition or can be left overnight to obtain relaxed shells with homeotropic anchoring as an initial condition. To induce the homeotropic to planar anchoring transition, these homeotropic shells are pipetted into another solution with only 1\% wt PVA and 0.1 M NaCl, without SDS. After introducing shells to the appropriate outer aqueous phase for the desired anchoring transition, the sample vial is gently mixed before pipetting into a viewing chamber (Grace Bio-Labs). An upright microscope in transmission mode fitted with crossed polarizers (Zeiss AxioImager M1m) and a high-resolution color camera (Zeiss AxioCam HRc) is used to capture polarized micrographs of shells.

Sodium chloride (NaCl, Fisher Scientific) is added to aqueous phases, ranging in concentration from 0.1 to 1 M, in order to increase the SDS interfacial density \cite{abbsurf}. In some samples, a higher concentration of NaCl in the inner aqueous phase compared to the outer phase can trigger osmotic swelling, in which water permeates through the CLC shell towards the inner aqueous droplet, increasing the inner droplet volume and thinning the shell over time \cite{osswell,alex-waltz,TranPRX}. However, we found the osmotic swelling of CLC shells to occur over tens of hours, while stripe transformations from the introduction or removal of SDS occur over a span of tens of minutes. Therefore, we do not expect osmotic swelling to have an effect on the initial pattern formation that is the focus of this work. Indeed, the time and the manner of pattern formation for shells with and without osmotic swelling is comparable.


\subsection{Landau-de Gennes modeling}

The numerical simulations model the CLC molecular orientation via the Landau-de Gennes free energy where a symmetric, traceless $3 \times 3$ matrix $Q_{\alpha \beta}(\mathbf{x})$  represents the nematic orientation order parameter and is defined at each site $\mathbf{x}$ of a cubic lattice, with indices $\alpha,\beta=x,y,z$ indicating the Cartesian directions \cite{ravnik,sussman}.  The 
free energy in the bulk regions of the liquid crystal reads
\begin{equation}
f_{\mathrm{bulk}}= \frac{A}{2}\, Q_{\alpha \beta}Q_{\beta \alpha} + \frac{B}{2} \,Q_{\alpha \beta} Q_{\beta \gamma} Q_{\gamma \alpha} + \frac{C}{4} ( Q_{\alpha \beta}Q_{\beta \alpha})^2,
\end{equation}
with summation over repeated indices implied. The constants $A$, $B$, and $C$ are set by the thermodynamic properties of the liquid crystal. Note that the scalar order parameter is given in terms of these parameters as $S_0 = (-B+ \sqrt{B^2-24AC})/6C$.   We also have an elastic component in the bulk which incorporates the energy penalties 
associated with splay, twist, and bend distortions:
\begin{align}
f_{\mathrm{grad.}} & = \frac{L_1}{2}\,( \epsilon_{\alpha \gamma \delta} \partial_{\gamma} Q_{\delta \beta}+2 q_0 Q_{\alpha \beta})^2 + \frac{L_2}{2} ( \partial_{\beta} Q_{\beta \alpha})^2 \nonumber \\
&  {} + \frac{L_{24}}{2} (\partial_{\alpha} Q_{\beta \gamma} \partial_{\gamma} Q_{\alpha \beta} - \partial_{\alpha} Q_{\alpha \beta} \partial_{\gamma} Q_{\beta \gamma}),  \label{eq:LdGFT}
\end{align}
where $\epsilon_{\alpha \beta \gamma}$ is the Levi-Cevita symbol and $\partial_{\alpha} \equiv \frac{\partial}{\partial x_{\alpha}}$ are the partial derivatives along the three spatial directions. When we square the two terms in the first line of Eq.~\eqref{eq:LdGFT}, we are left with a sum over the remaining free indices $\alpha$ and $\beta$. More details are given in, \textit{e.g.}, \cite{fukuda}. We include a splay-bend coefficient $L_{24}$ that is a total derivative term, but will be important in our system since we will be interested in interfacial phenomena.

At interfaces, we also have an anchoring energy that we model using a Rapini-Papoular surface potential \cite{rapinipapoular}. For homeotropic anchoring strength $W_0$ and planar anchoring strength $W_1$, 
the total interface free energy reads:
\begin{align}
f_{\mathrm{in.}} & =  \int \mathrm{d} A \Big\{ W_0 (Q_{\alpha \beta}-Q^{\parallel}_{\alpha \beta})^2  \nonumber \\ & {} +W_1 [(\bar{Q}_{\alpha \beta}-\bar{Q}^{\perp}_{\alpha \beta})^2 + (Q_{\alpha \beta}Q_{\alpha \beta}-3S_0^2/2)^2\Big\} ,
\end{align}
where $Q^{\parallel}_{\alpha \beta} = 3 S_0 (\hat{s}_{\alpha} \hat{s}_{\beta}- \delta_{\alpha \beta}/2)$ is the orientation tensor
 constructed from the interface's surface normal vector $\hat{\mathbf{s}}$ and where $\bar{Q}_{\alpha \beta}=Q_{\alpha \beta}+S_0 \delta_{\alpha \beta}/2$. The tensor $\bar{Q}^{\perp}_{\alpha \beta}$ is the projection of $\bar{Q}_{\alpha \beta}$ onto the plane of the interface, so $\bar{Q}^{\perp}_{\alpha \beta}=(\delta_{\alpha \beta} - s_{\alpha} s_{\delta})\bar{Q}_{\delta \gamma}(\delta_{\gamma \beta}-s_{\gamma}s_{\beta})$.
 
 This free energy is then minimized over a computational domain using a conjugate gradient method in the ALGLIB package (\url{http://www.alglib.net/}).  Note that the minimization here does not reflect the actual dynamics of a liquid crystalline system. In reality, the free energy would be minimized in a manner consistent with liquid crystal hydrodynamics. Nevertheless, we can obtain a qualitative idea about the evolution of various patterns by monitoring the states of the system during the minimization procedure from some specified initial condition while tracking any long-lived transient states. Often, the minimization procedure will end within local free energy minima (metastable states) which are also of interest.
 
 In the experiments presented here, we consider 5CB mixed with a chiral dopant as our cholesteric liquid crystal. Therefore, we make use of some standard values for this compound in our simulations \cite{ravnik}.   We set $A=-1$, as this can be set simply by choice of free energy units. We then utilized a one-constant approximation   with $B=-12.33$, $C=10.06$, and $L_1=L_2=2.32$ (Fig.~\ref{fig:unwinding}), consistent with the parameters of 5CB and a lattice spacing $\Delta x \approx 4.5~\mathrm{nm}$ close to the nematic correlation length (size of nematic defect core).  These particular choices were not important for determining the qualitative features of the simulation, including the formation of the undulation instability that is the focus of this work.  For the other shell simulations in  Figs.~\ref{fig:arches} and \ref{fig:HomToPla}, we used the two-constant approximation with $B=-1.091$, $C=0.6016$, $L_1=0.003805$, and $L_2=2L_{24}=0.01141$. These latter two constants were chosen to correspond to $K_1=K_3 \approx 10~\mathrm{pN}$ and $K_2 = 5~\mathrm{pN}$ for the characteristic splay, bend, and twist elastic constants, respectively, for 5CB \cite{elastic5CB}. The choice of constants yields a large lattice spacing $\Delta x \approx 30~\mathrm{nm}$. Note that the bulk free energy constants $B$ and $C$ are not so important in this case because we simulate deep in the nematic phase (low temperature) with a large $\Delta x$ so that all the lattice sites retain the equilibrium value $S_0$ of the scalar order parameter. The minimization procedure essentially only minimizes the elastic component as the bulk remains at a uniaxial minimum.  For the simulations with a deformable CLC shell interface in Fig.~\ref{fig:freeinterface}, we tuned the cholesteric to an isotropic-nematic phase transition by setting $A=5.284$, $B=-11.2612$, and $C=0.88889$, with the elastic constants $L_1 = L_2 = 2L_{24}= 0.928$. The various corresponding dimensions of the shell are given in the figure captions  in SI units.


\section{Results \label{sec:Results}}

 We now examine the undulation instability of CLC stripes on shells for both homeotropic and planar anchoring transitions, using experiments and simulations. At one end of an anchoring transition, the equilibrated \textit{planar} state is absent of stripes, with director defects adding up to a topologically-required index of $+2$ \cite{alex-defects, alex-waltz}. An example planar shell with four +1/2 defect points is shown in Fig.~\ref{fig:FCDwinding}-i. In the planar state, the pitch axis is oriented radially, with the cholesteric layers forming concentric shells. At the other end of an anchoring transition, the \textit{homeotropic} state of a CLC shell typically stabilizes focal conic domains, seen as double spirals at the shell surface, in which the cholesteric twists along the shell interface (top row of Fig.~\ref{fig:FCDunwind} and Fig.~\ref{fig:surface}-i). The pitch axis in the homeotropic state is oriented parallel to the surface. When the CLC shell is left to equilibrate for a sufficiently long time, the stripes eventually arrange into lines of latitude on the shell, with stripes terminating at two focal conic domains, as forced in by the spherical topology \cite{TranPRX}. Converting between these two states requires tilting the pitch axis orientation with respect to the surface, yielding undulating stripes that are featured in the bottom row of Fig.~\ref{fig:FCDunwind} and are the subjects of interest in this paper. 

\subsection{Planar to homeotropic transitions}

\begin{figure}[htp]
\centering
\includegraphics[width=0.4\textwidth]{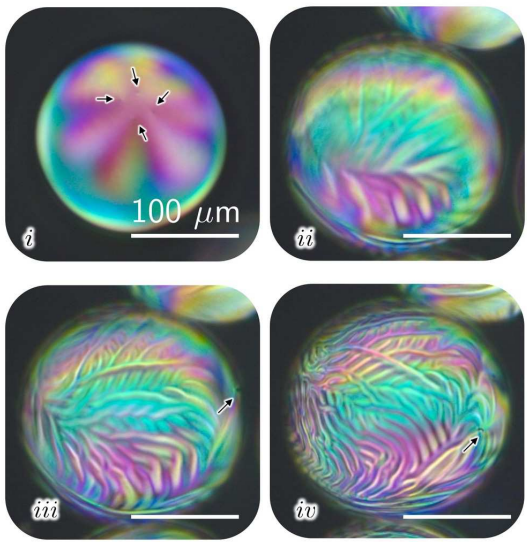}
\caption{\label{fig:FCDwinding}  An initially planar-anchored CLC shell (a) has 10 mM SDS added to the outer aqueous phase, detailed in Methods (Sec.~\ref{sec:setup}). As the surfactant adsorbs to the shell interface and interacts with the CLC, it begins to orient the CLC molecules perpendicular to the interface, tilting the pitch axis away from the radial direction. The change in anchoring generates large stripes with a periodicity around twice the pitch, analogous to those seen in Fig.~\ref{fig:FCDunwind} in the bottom panels. Arrows point out director defects. Video can be viewed at \cite{nikonsmallworld}.} 
\end{figure}

We begin first with the planar to homeotropic anchoring transition, featured in Fig.~\ref{fig:FCDwinding}. Homeotropic anchoring is induced on the outer shell surface experimentally through the addition of a surfactant (SDS, see Sec.~\ref{sec:setup}) to the outer aqueous phase. The onset of the stripe instability is shown in Fig.~\ref{fig:FCDwinding}. Note that the undulated stripes do not conform to a discernible pattern, but run in different directions along the shell surface. The nematic defects in the planar to homeotropic transition do not appear to play a large role in ordering the undulating stripes, although the stripes can be observed terminating at the director defects, highlighted in Fig.~\ref{fig:FCDwinding}-iv.

  \begin{figure}[htp]
\centering
\includegraphics[width=0.43\textwidth]{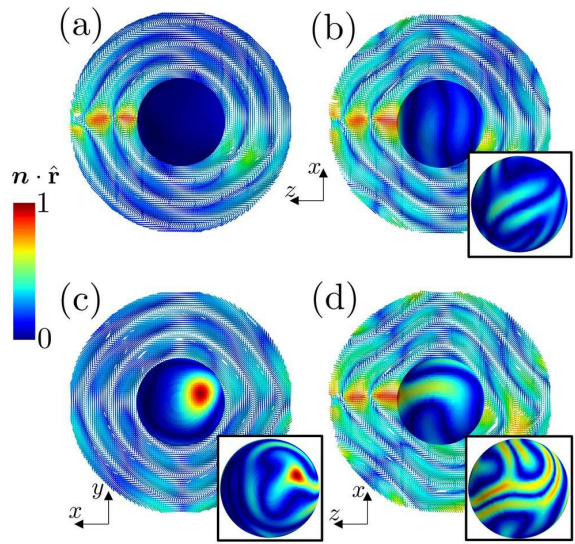}
\caption{\label{fig:arches} An initially planar shell (a) with a $6.6~\mu\mathrm{m}$ diameter and a $2.1~\mu\mathrm{m}$ thickness is minimized in the presence of moderate homeotropic boundary conditions ($W_0=0.04$ corresponding to about $2 \times 10^{-4}~\mathrm{J}/\mathrm{m}^2$). The cholesteric has a $1.2~\mu\mathrm{m}$ pitch. After $t=5000$ minimization steps, the concentric cholesteric layers begin to undulate and bend toward the surface, generating stripes shown in (b). The inset is the outer surface. A side view of (b) is plotted in (c). The planar defect is marked in red by the $\mathbf{n} \cdot \hat{\mathbf{r}}$ color map. As the system evolves, this region becomes increasingly homeotropic at both the inner and outer surfaces. Eventually, the bent, undulating CLC layers form the focal conic domains characteristic of the homeotropically-anchored state. We can see the beginning of these characteristic stripes after $t=60000$ minimization steps in (d). }
\end{figure}

We turn to simulations to elucidate the bulk CLC layer arrangements that bring about the undulating stripe patterns. The stripe behavior is simulated by taking a planar CLC shell configuration as the initial condition [Fig.~\ref{fig:arches}(a), left] and minimizing the free energy under homeotropic anchoring conditions. Large stripes are spawned at early stages of the transition [Fig.~\ref{fig:arches}(b)], reminiscent of those seen in experiments. Cross sections of the shell reveal the source of the large stripes to be undulation of the concentric CLC layers, evident from comparing the cross sections in Fig.~\ref{fig:arches}(a) and (b). During the minimization, homeotropic anchoring increases most significantly at the director defect, the area of highest distortion [red region in Fig.~\ref{fig:arches}(c)]. Interestingly, far from the director defect, the CLC layers are not as undulated as they are near the defect, evident in the cross section of Fig.~\ref{fig:arches}(c). That undulations are most pronounced near the director defect suggests that pitch axis reorientation occurs first at these local regions of disorder before propagating to the rest of the system. Eventually, the typical stripe pattern of homeotropic CLC shells develops on the surface, forming focal conic domains and indicating that the pitch axis completes its reorientation and becomes parallel to the shell surfaces. The simulation results further show that the distortions due to the changing anchoring are confined to regions closest to the shell interfaces, with the most bent and distorted cholesteric layers near the surface [Fig.~\ref{fig:arches}(d)].

\subsection{Homeotropic to planar transitions}

\begin{figure*}[htp]
\centering
\includegraphics[width=0.93\textwidth]{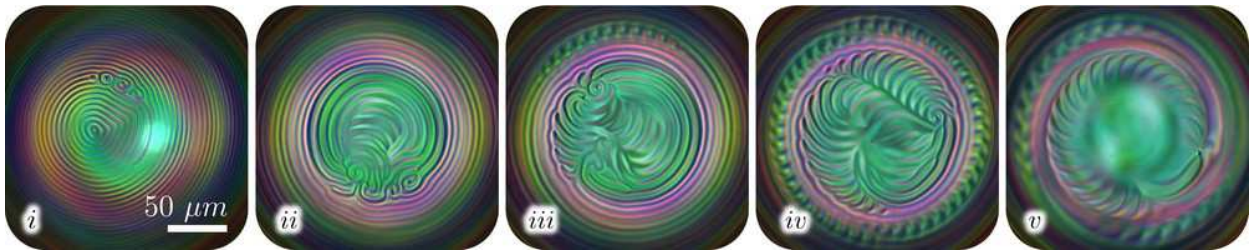}
\caption{\label{fig:PRXunwind} A homeotropic CLC shell left to equilibrate for one month has two focal conic domains at its poles,  with additional defects in the thinnest region of the shell (i). During the homeotropic to planar anchoring transition, the focal conic domains unwind and widen to create greater planar anchoring at the shell surface. As the transition continues, stripes begin to locally undulate with a $2P_0$ periodicity (iii-iv), where $P_0 = 5$ $\mu$m. At the end of the transition the planar state is reached, with remnant undulations on the shell. Reproduced from Tran \textit{et. al.}, Supplemental Video 6 \cite{TranPRX}. }
\end{figure*}

A similar stripe instability also occurs for a homeotropic to planar anchoring transition. However, in contrast to the reverse transition, the onset of the stripe instability induced by imposing planar anchoring appears to be sensitive to the initial configuration of the system. We examine first the transition for an equilibrated, homeotropic CLC shell before turning to the transition for metastable, homeotropic states with the characteristic focal-conic-like domains decorating the CLC shell surface.

A CLC shell left to equilibrate in a solution with 7 mM SDS, 1\% wt PVA, and 0.1 M NaCl for a month is shown in Fig.~\ref{fig:PRXunwind}(i), first presented in \cite{TranPRX}. It has two focal conic domains at it poles, with the pole in the thinnest region of the shell highlighted in Fig.~\ref{fig:PRXunwind}(i). Transferring the shell to an aqueous solution without SDS, results in slow removal of surfactant from the interface and a reduction in the homeotropic anchoring strength. The double spirals of the polar focal conic domains unwind first, resulting in stripes of a larger periodicity at the poles, seen in Fig.~\ref{fig:PRXunwind}(ii). As the stripes unwind, they also widen to accommodate greater regions of planar anchoring at the interface. At some point during the anchoring transition, stripes near the thinnest region of the shell begin to undulate with a periodicity of twice the pitch, shown in Fig.~\ref{fig:PRXunwind}(iii) \& (iv). The undulations extend along the stripes as the transition continues. Near the end of the transition to planar anchoring, most of the stripes have been removed from the system, with a few undulations remaining. One of the excess defects near the focal conic domain in Fig.~\ref{fig:PRXunwind}(i) becomes the topologically required director defect in Fig.~\ref{fig:PRXunwind}(v).

\begin{figure}[htp!]
\centering
\includegraphics[width=0.43\textwidth]{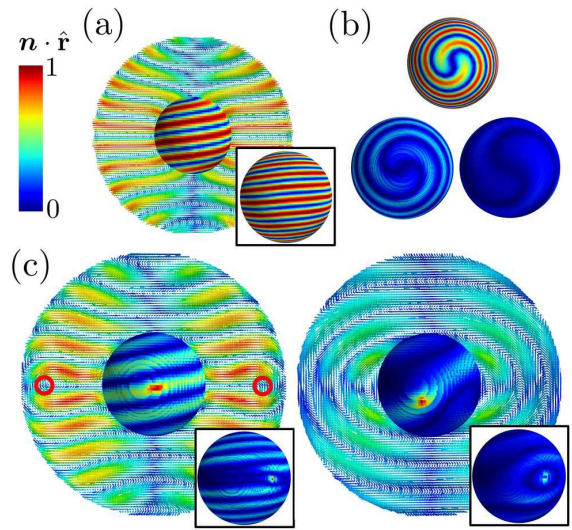}
\caption{\label{fig:HomToPla} A shell at equilibrium with homeotropic anchoring in (a) is relaxed under planar anchoring conditions ($6.6~\mu\mathrm{m}$ diameter, $2.1~\mu\mathrm{m}$ thickness, and $W_1=0.004$ corresponding to about $2 \times 10^{-5}~\mathrm{J}/\mathrm{m}^2$). Top views of the shell during the anchoring transition to planar are plotted in (b). The double spiral unwinds as the system approaches its new energetically-favorable, planar-anchored state. Cross sections of the shells over the course of the transition, from $t=2000$ (left) and $t=10000$ (right) minimization steps, are plotted in (c). At $t=2000$, the layers start to undulate in the shell interior (seen most prominently in the red regions). At the equator, the layer undulates such that a layer pinch-off gives rise to two pitch defects (red circles) that eventually become the planar defects at the end of the transition. At $t=10000$, the undulated layers merge and become concentric. Eventually, a planar state is achieved, similar to Fig.~\ref{fig:arches}(a) (left).}
\end{figure}

The absence of undulating stripes during the majority of the anchoring transition is captured through simulations, plotted in Fig.~\ref{fig:HomToPla}, where an equilibrated, homeotropically-anchored cholesteric shell relaxes to a planar-anchored state. The starting condition, shown in Fig.~\ref{fig:HomToPla}(a), has two characteristic double-spirals at both poles that correspond to the topologically-required, focal-conic-like domains where the pitch axis has a defect.  During relaxation, these double spirals unwind as the pitch axis tilts toward the shell interior. We see a top view of the unwinding in Fig.~\ref{fig:HomToPla}(b). Meanwhile, the layers on the interior of the shell begin to undulate, shown on the left of Fig.~\ref{fig:HomToPla}(c). The layers near the equator undulate in opposite directions and pinch off, creating another pitch defect that eventually becomes the topologically-required nematic defect in the planar-anchored shell.  Such a mechanism was conjectured in \cite{TranPRX}, but is observed here explicitly in the simulation. Eventually, the pitch axis points radially, and we obtain a state close to the equilibrated planar configuration, seen on the right of Fig.~\ref{fig:HomToPla}(c). We can discern from the insets in Fig.~\ref{fig:HomToPla}(c) that one of the nematic defects (red spot on the outer surface) sits near the shell equator. Faint stripes are still observable near the director defect, shown through the inset of Fig.~\ref{fig:HomToPla}(c), right. Generally, a secondary stripe instability does not materialize, reminiscent of the experiment featured in Fig.~\ref{fig:PRXunwind}, where stripes are capable of becoming wider and more planar without undulating in the majority of the system.

\begin{figure}[htp]
\centering
\includegraphics[width=0.44\textwidth]{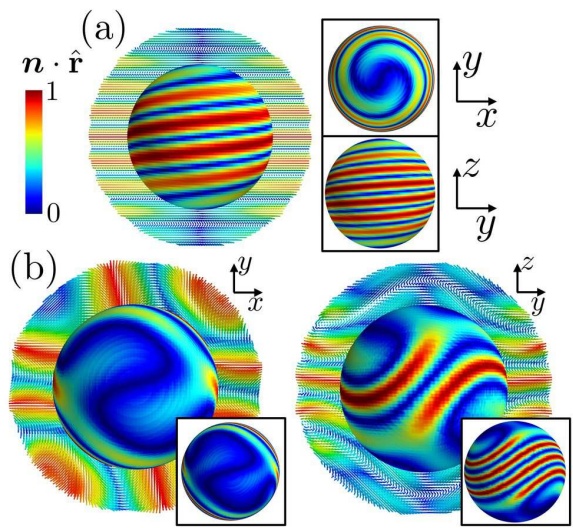}
\caption{\label{fig:unwinding} A CLC shell simulation ($0.84~\mu\mathrm{m}$ diameter and $0.18~\mu\mathrm{m}$ thickness using the one-constant approximation, see Sec.~\ref{sec:setup}), with a color map of the radial component $\mathbf{n} \cdot \hat{\mathbf{r}}$ of the director for a cholesteric phase with a $0.18~\mu\mathrm{m}$ pitch. The insets show the outer surfaces. Here, an initially $\hat{z}$-axis oriented pitch axis configuration (a) is set to relax with planar anchoring of magnitude $W_1=0.25$, which corresponds to a strength of $2 \times 10^{-4}~\mathrm{J}/\mathrm{m}^2$.  The minimization procedure finds a local minimum in the free energy corresponding to the partially unwound stripes shown in (b), with a top view on the left and a side view on the right. The poles develop undulated stripes, while the equatorial stripes partially unwind. }
\end{figure}

However, it is possible to observe the undulated stripe instability in simulations by using a thinner, smaller shell and slightly altering the initial equilibrium, homeotropic configuration, as shown in Fig.~\ref{fig:unwinding}.   Using a uniform CLC ground state with a vertical pitch axis as the initial shell configuration creates distortion of the polar focal conic domains, shown on the left of Fig.~\ref{fig:unwinding}(a). The distorted focal conic domains are regions  where deformation of the CLC is easiest, observed also in our prior work \cite{TranPRX}. With tilting of the pitch axis beginning at these regions near the poles, the centers of the stretched double spirals are also where the undulations first appear. Under the anchoring transition, the stretched double spirals unwind and widen only in the double spiral region, leaving behind an undulated, planar stripe, shown in Fig.~\ref{fig:unwinding}(b). The simulation is then trapped in this configuration, indicating a metastable state. Cross sections on the right of Fig.~\ref{fig:unwinding}(b) reveal that the undulated stripes are a direct result of CLC layer undulations. The left panel of Fig.~\ref{fig:unwinding}(b) further reveals that the stripe undulation appears on both the inner and outer surfaces of the shell. These modulations of the director, perpendicular to the initial stripes, mimic what is seen at the former focal conic domain of the experiment featured in Fig.~\ref{fig:PRXunwind}i-ii \cite{TranPRX}. As the shell adapts to planar anchoring in Fig.~\ref{fig:PRXunwind}, the double spiral region also develops undulated stripes, matching the simulated behavior of the distorted focal conic domains in Fig.~\ref{fig:unwinding}. 

\begin{figure}[htp]
\centering
\includegraphics[width=0.42\textwidth]{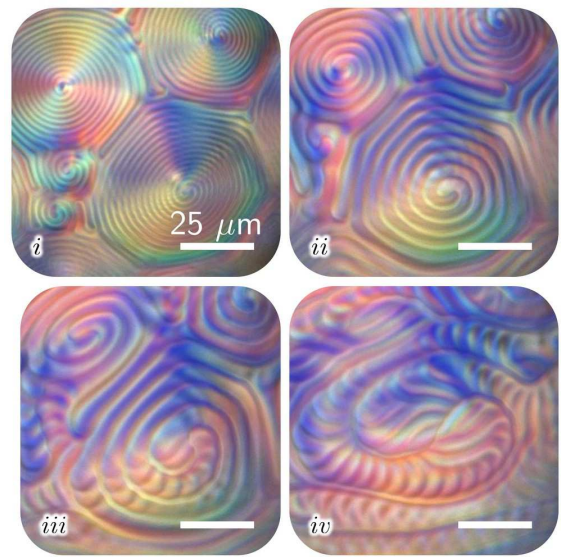}
\caption{\label{fig:surface} A thick cholesteric shell in an aqueous solution with 7 mM SDS, 1\% wt PVA, and 0.1 M NaCl has a focal conic domain texture initially. The pitch is 5 $\mu$m. The shell is transferred to another, similar aqueous solution, but without SDS, and the evolution of the texture is observed (i-iv). As the outer interface loses homeotropic anchoring strength with surfactant removal to the surrounding solution, the planar anchoring stripes widen (i-ii). When the stripes widen to $\sim 2 P_0 \approx 10$ $\mu$m, the stripes fill with perpendicular stripes of a second periodicity also $\sim 2 P_0$. See Video 1 in Supplemental Material. } 
\end{figure}

Curiously, undulating stripes in the homeotropic to planar anchoring transition are more pervasive in experimental systems where the initial homeotropic state is meta-stable and have many focal conic domains, exemplified in Fig.~\ref{fig:FCDunwind}-i and Fig.\ref{fig:surface}-i. As the surfactant is washed off, the stripes widen to accommodate larger regions of planar anchoring. When the stripes reach a width of approximately twice the pitch $P_0$, the stripe interior becomes filled in with secondary, arch-like stripes of a periodicity $\sim 2P_0$ that run along primary stripes, covering the entire shell surface (Fig.~\ref{fig:FCDunwind}-iii-iv). Primary stripes being filled in with undulating, secondary stripes also matches the simulation results of Fig.~\ref{fig:unwinding}. However, it remains unclear why stripes in metastable configurations more readily undulate than those in equilibrium configurations. The initial, seeded planar anchoring at the surface of the simulation in Fig.~\ref{fig:unwinding} allows for the undulated stripes to emerge during the anchoring transition. This implies that larger, local regions of planar anchoring may be necessary in experiments for secondary, undulated stripes to appear. This could possibly occur in experiments from the segregation of surfactants into regions of favorable anchoring, reported in \cite{lt-sciadv}. Shell thickness heterogeneity in experiments may additionally play a role, as the local undulations of Fig.~\ref{fig:PRXunwind} occur near the thinnest region of the shell. Indeed, the simulated shell of Fig.~\ref{fig:unwinding}, exhibiting pronounced, undulated stripes, is smaller and thinner than that of Fig.~\ref{fig:HomToPla}. 

Despite slight differences between homeotropic and planar anchoring transitions, the emergence of undulating stripes with changes in anchoring is a general phenomenon, demonstrated through the above experiments and simulations of CLC shells. Simulation results reveal that the stripe instability is a consequence of CLC layer undulation, resulting from the pitch axis tilting to accommodate the new anchoring conditions. In the following section, we formulate a simple model through the lens of the Helfrich-Hurault instability to describe how incompatible anchoring can trigger undulations in the cholesteric bulk.


\section{Analysis\label{sec:Analysis}}

\begin{figure}[htp]
\centering
\includegraphics[width=0.48\textwidth]{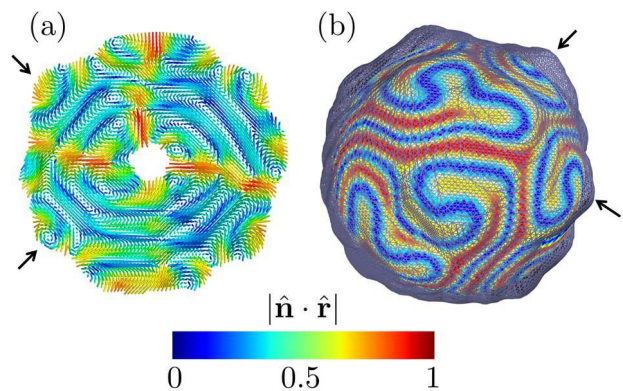}
\caption{\label{fig:freeinterface} (a) Cross section through a thick cholesteric shell with a free interface from contact with an isotropic phase, illustrating the distortions of the cholesteric layers near the surface due to the incompatible homeotropic anchoring. In the bulk, the layers settle into concentric shells. Arrows indicate the focal conic ``hills.'' The color indicates the director $\hat{\mathbf{n}}$ orientation relative to the radial direction $\hat{\mathbf{r}}$. The pitch $P_0$ to shell thickness $t$ ratio here is $t/P_0 \approx 2$. (b) The rough interface and the director distribution just underneath the surface is plotted. Note that the bumps are correlated with the double spirals  of the focal conic domains (arrows).    }
\end{figure}

A key feature of CLCs is that anchoring changes are confined to a small boundary layer $\ell$ near the interface, while retaining the overall layer structure of the cholesteric bulk, seen in the simulation results of Fig.~\ref{fig:arches}. This is true even if the interface is allowed to deform, as would happen at the cholesteric-isotropic interface, simulated in Fig.~\ref{fig:freeinterface}.  Here, a shell with concentrically-arranged cholesteric layers deforms the cholesteric-isotropic interfaces of the shell, with the layers bending most significantly near the interfaces. In larger systems, the bulk layers generally remain in their former configuration. A similar boundary layer may be found in cholesterics confined to hybrid-anchored cells \cite{archtexture}.  

 The instabilities considered here result from cholesteric layer strain imposed by a change in anchoring conditions. The strain is relieved via a periodic modulation of the layers, which manifests in the stripe instability described in Sec.~\ref{sec:Results}.  This phenomenon is reminiscent of the classic Helfrich-Hurault mechanism, in which cholesteric layers are strained via an applied magnetic field \cite{helfrich,hurault1}. We develop a similar analysis of the instability observed during anchoring transitions. 
 
 We begin by considering the Frank free energy density of the cholesteric:
\begin{eqnarray}
f_{\mathrm{n}}  &=& \frac{K_1}{2}[\hat{\mathbf{n}}(\nabla \cdot\hat{\mathbf{n}} ) ]^2+\frac{K_2}{2}[ \hat{\mathbf{n}} \cdot(\nabla \times \hat{\mathbf{n}})+q_0]^2 \nonumber   \\
&&\quad + \frac{K_3}{2}[(\hat{\mathbf{n}} \cdot \nabla) \hat{\mathbf{n}}]^2   \label{eq:frank}
\end{eqnarray}
where $K_1$, $K_2$, and $K_3$   are the splay, twist, and bend elastic constants, respectively. This energy does not properly take into account defects in the nematic director, but it can describe the cholesteric layer structure.  Minimization of the $K_2$ term yields the usual cholesteric ground state, characterized by a spiraling nematic director $\hat{\mathbf{n}}= \sin(\mathbf{q}\cdot \mathbf{x}-q_0u) \hat{\mathbf{q}}^1_{\perp}+\cos(\mathbf{q}\cdot \mathbf{x}-q_0u)\hat{\mathbf{q}}^2_{\perp}$, with $\hat{\mathbf{q}}$ being the pitch direction and $\hat{\mathbf{q}}^{1,2}_{\perp}$ being two orthogonal basis vectors spanning the subspace perpendicular to the pitch. We also have $|\mathbf{q}|=q_0=2 \pi/P_0$, with $P_0$ being the pitch. The phases $q_0 u$ are the distortions of the cholesteric layers and $u=0$ for a perfect cholesteric helix.
The field $u(\mathbf{x})$ may be interpreted at large length scales as a displacement of the cholesteric layer spacings.

It can be shown that small deviations $u$ away from the cholesteric ground state have a free energy  analogous to a smectic (layered) liquid crystal \cite{dgennes-prost,RadzihovskyLubensky}:
\begin{equation}
F_e=\int \mathrm{d}^3\mathbf{x}\left[\frac{\bar{B}}{2}
 \left(\frac{\partial u}{\partial z}-\frac{(\nabla_{\perp} u)^2}{2}\right)^2+\frac{\bar{K}}{2} \left( \nabla_{\perp}^2 u\right)^2 \right],
 \label{eq:layerE}
\end{equation}
where we integrate over the entire sample volume,  $\bar{K} \approx 3K_3/8$ is an effective layer bending modulus, and $\bar{B}=K_2q_0^2$ is an effective layer compression modulus  \cite{LubenskyHydro,dgennes-prost,RadzihovskyLubensky}. The gradient $\nabla_{\perp}$ represents  the derivatives perpendicular to the layer orientation. The free energy in Eq.~\eqref{eq:layerE} is consistent with a rotational invariance of the overall layer structure \cite{kleman-odl}.  

In our experiments, we always apply an anchoring that favors a pitch axis tilting, either away from the interface normal for planar to homeotropic transitions, or toward the interface normal for homeotropic to planar transitions. In both cases, such an anchoring will introduce the following free energy contribution at the interface for small tilts $u$:
\begin{equation}
F_{\mathrm{a}}=-\frac{|W|}{4}\int \mathrm{d}^2 \mathbf{x}\,(\nabla_{\perp}u)^2, \label{eq:anchoring}
\end{equation}
where we integrate over the interface surface and $W$ is the strength of the tilt-inducing anchoring \cite{odlanchoring}.  Much like an applied magnetic or electric field, this free energy contribution will  \textit{break} the rotational symmetry implicit in the layer free energy in Eq.~\eqref{eq:layerE}, driving an undulation instability.

 Let us now consider how the instability would work for a flat interface (\textit{\textit{i.e.}} a small patch of the CLC shell) located at the $z=\ell$ plane. We will assume that $\ell$ corresponds to the penetration depth of the deformations induced by the anchoring, so that there are no deformations $(u=0)$ at $z=0$.   The characteristic wave vector of the instability may be found by looking at the lowest harmonic of the $u$ field, which we assume corresponds to a modulation along the $x$-direction: $u(x,z)=u_0 \sin(\pi z/2\ell) \cos(q_x x)$, where $\ell$ is a characteristic length over which the deformation occurs within the CLC bulk.  For a CLC free interface, we expect the deformations to be confined to a length $\ell$ on the order of the cholesteric pitch or half-pitch \cite{freeCholAnchor}. Substituting in this ansatz and integrating Eq.~\eqref{eq:layerE} and \eqref{eq:anchoring}, we find $f_s$, a free energy per unit area of interface:
 \begin{eqnarray}
f_{\mathrm{s}}  &=& u_0^2 \left[\frac{ \pi ^2 \bar{B}}{32\ell}+\frac{\ell q_x^4}{512} \left(9 \bar{B}u_0^2+64 \bar{K}\right)-\frac{Wq_x^2 }{8} \right] .    \label{eq:HH}
\end{eqnarray}
We find that when  $W>W_c=\pi \sqrt{\bar{K}\bar{B}}$, the free energy in Eq.~\eqref{eq:HH} is minimized for a non-zero $u_0$ (\textit{i.e.}, an undulated state). Moreover, the critical wave vector associated with the modulation is $q_x^*=(\bar{B}/\bar{K})^{1/4}(\pi/2\ell)^{1/2}$. Putting in typical elastic constants for 5CB, and assuming a 5~$\mu$m pitch $P_0$ (setting $\ell=P_0$ equal to the pitch), we expect $W_c \approx10^{-5}~\mathrm{J}/\mathrm{m}^2 $ and a modulation wavelength $\lambda_x^*  \equiv 2\pi/q_x^*\approx 2P_0$. This is a relatively weak anchoring strength, so we do generically expect to see a modulation under the experimental conditions presented here. 

A modulation wavelength of about twice the pitch is also consistent with our observations, detailed previously in Sec.~\ref{sec:Results}. However, we note that this spacing can be larger than $2P_0$, especially when the initial configuration of the layers does not correspond to a uniform cholesteric helix running either along or perpendicular to the interface. Indeed, aspects not captured by this simple argument are other, more complex arrangements of the cholesteric layers with initially homeotropic anchoring. For instance, the cholesteric layers can bend to form u-shapes near the interface, resulting in a distribution of $\lambda^{\pm}$ defects \cite{TranPRX,freeCholAnchor}, leading to a stripe periodicity equal to the pitch, instead of a half-pitch as one would obtain for a uniform, undistorted helix oriented parallel to the interface. 


\section{Discussion \& Conclusion \label{sec:Conclusion}}

 We have now shown how undulation instabilities develop at free CLC interfaces when the anchoring changes from homeotropic to planar or vice-versa. The instability is driven by a strain in the layers due to the reorientation of the pitch axis near the interface. One typically finds that the modulations have a periodicity equal to twice the cholesteric pitch. We have shown that this instability is analogous to the Helfrich-Hurault mechanism, with the anchoring change playing the role of an ``applied field''.  A basic argument yields a reasonable estimate of the undulation instability periodicity and the critical anchoring strength at which we might expect to see the instability.

 Confinement of the CLC within a spherical geometry necessitates the presence of defects that appear to influence the conformation of the stripe instabilities, with secondary stripes terminating at director defects for homeotropic transitions and extending along primary stripes for planar transitions. For the homeotropic anchoring transition, the initial director defects serve as favorable sites for cholesteric layer rearrangements and undulations. For the planar anchoring transition, the focal conic domains, pitch defects, also act as locations of initial pitch axis reorientation, evident from stripes becoming wider at double spirals first. We note that although the topologically-required defects serve as regions of easy deformation in the system, the presence of defects is not necessary for the onset of undulations. Indeed, the Helfrich-Hurault instability requires only a local geometric incompatibility, instead of a global frustration, for undulations to occur. The type of defect, whether director or pitch, merely reflects the cholesteric layer orientation with respect to the confining boundaries.

We hypothesize that the difference in the appearance and ordering of the stripe instability in homeotropic versus planar transitions lies in the shortest path for the pitch axis reorientation. For the homeotropic transition, the pitch axis is initially radial. To conform to the new homeotropic anchoring condition, the pitch axis must tilt to become tangent to the surface. However, every tilt direction from radial is equivalent. The disorder of the stripe instability for the homeotropic transition then lies in the degeneracy of the pitch-axis tilt direction. On the other hand, for the planar transition, the pitch axis is initially tangent to the spherical surface. Therefore, tilting the pitch axis \textit{along} the direction of its initial orientation is the shortest path to pointing radially. This constraint prescribes a set direction for undulations to take place, along the initial (primary) stripes on the surface. We additionally note that the secondary, undulated stripes in the planar transition often have an arching shape in experiments, reminiscent of Bouligand arches that emerge when a surface cuts a CLC at an angle to the pitch axis \cite{boulig-biochol, bouligarches, boulig-spherulites}. The exact connection between the stripe instability and Bouligand arches remains to be further explored.
  
It would be interesting to develop a more faithful simulation that takes into account the liquid crystal dynamics more properly. Our free energy minimization procedure assumed an over-simplified relaxation mechanism for the nematic director. In the real system, hydrodynamic effects may be important. Moreover, simulating a true free interface could also shed light on the role of the interface in triggering or stabilizing the stripe instability. We are currently limited to looking at interfaces between a cholesteric and isotropic phase, which has a fixed, weak homeotropic anchoring for the cholesteric. Furthermore, a description that takes into account heterogenous shell thicknesses and distributions of anchoring-inducing, surface-active agents may also be needed to fully capture the states observed in experiments.

Another unexplored aspect is the relationship of this instability to the stripe instability observed in \textit{nematic} liquid crystals under hybrid anchoring \cite{han-str-1,han-str-2,han-str-3,han-str-4,han-str-5}. Our system is similar when we transition from a planar to a homeotropic shell configuration because most of the bulk cholesteric layers maintain a nematic director orientation parallel to the shell surface (\textit{i.e.}, a planar orientation) while the outer layers have homeotropic anchoring. A hybrid-anchored cholesteric is even more complicated than the nematic case, due to the interplay between the anchoring and the cholesteric twist \cite{dozov}. In the case of the hybrid-anchored nematic, boundary terms in the  elastic free energy (\textit{e.g.}, the saddle splay) play an important role in determining the onset of modulations \cite{hybridK24,hybridK13}.  Although we include such terms in our free energy, we do not study the effect of this term systematically. We do expect such terms to contribute to our CLC shell, as we also have a boundary-driven instability. 

We have established how frustration between the surface and bulk ordering of a chiral material can drive an undulating instability, generating stripe patterns with a periodicity larger than that of the material itself. The Helfrich-Hurault model was expanded beyond applied fields and mechanical strains to encompass also changes in the surface anchoring as a source of the instability. This work lays the foundation for further study of the Helfrich-Hurault model in systems where boundary conditions can be freely adjusted and curved. 
\qquad

\begin{acknowledgments}
We thank R. D. Kamien, F. Livolant, and T. Lopez-Leon for useful discussions. Computational support was provided by the University of Tennessee and Oak Ridge National Laboratory's Joint Institute for Computational Sciences. M.O.L. gratefully acknowledges partial funding from the Neutron Sciences Directorate (Oak Ridge National Laboratory), sponsored by the U.S. Department of Energy, Office of Basic Energy Sciences. L.T. acknowledges support from the Simons Society of Fellows of the Simons Foundation.
\end{acknowledgments}


\end{document}